\documentstyle[eqsecnum,pre,preprint,aps,epsfig]{revtex}

\tightenlines

\newcommand{\bq}{\begin{equation}}
\newcommand{\eq}{\end{equation}}
\newcommand{\bqa}{\begin{eqnarray}}
\newcommand{\eqa}{\end{eqnarray}}
\newcommand{\ben}{\begin{enumerate}}
\newcommand{\een}{\end{enumerate}}
\newcommand{\bc}{\begin{center}}
\newcommand{\ec}{\end{center}}
\newcommand{\bqb}{\begin{eqnarray*}}
\newcommand{\eqb}{\end{eqnarray*}}

\begin{document}

\draft
\preprint{corrected version}

\title{\vspace{2cm}  A high energy determination of 
Yukawa couplings\\ in SUSY models
\footnote{Partially supported by EU contract HPRN-CT-2000-00149}}
\author{M. Beccaria$^{a,b}$,
F.M. Renard$^c$ and C. Verzegnassi$^{d, e}$ \\
\vspace{0.4cm}
}

\address{
$^a$Dipartimento di Fisica, Universit\`a di
Lecce \\
Via Arnesano, 73100 Lecce, Italy.\\
\vspace{0.2cm}
$^b$INFN, Sezione di Lecce\\
\vspace{0.2cm}
$^c$ Physique
Math\'{e}matique et Th\'{e}orique, UMR 5825\\
Universit\'{e} Montpellier
II,  F-34095 Montpellier Cedex 5.\hspace{2.2cm}\\
\vspace{0.2cm}
$^d$
Dipartimento di Fisica Teorica, Universit\`a di Trieste, \\
Strada Costiera
 14, Miramare (Trieste) \\
\vspace{0.2cm}
$^e$ INFN, Sezione di Trieste\\
}

\maketitle

\begin{abstract}

We consider the production, at future lepton colliders, of final 
fermion, sfermion, scalar pairs in SUSY models. For third family 
fermions and sfermions and for charged Higgses, the leading 
Yukawa effect at one loop for large c.m. energies
comes from a linear logarithm of Sudakov type, that only 
depends, in the MSSM, on one SUSY mass scale and on 
$ \tan \beta $. Assuming a relatively light SUSY scenario, 
we illustrate a possible determination of $\tan \beta $ at 
c.m. energies of about 1 TeV, working
systematically at subleading logarithmic accuracy, at the 
one-loop level.

\end{abstract}

\section {Introduction}
The possibility that virtual electroweak effects of 
Supersymmetric models at
future lepton colliders\cite{LC,CLIC} can be described,
for sufficiently large c.m. energies, 
by a logarithmic expansion of Sudakov type, has been examined 
in recent papers \cite{susylog,tgb,scalar}
in the simplest case of the MSSM. The first considered process 
has been that of final fermion pair production. This has been 
only treated at one-loop, computing 
for both massless \cite{susylog} and massive
\cite{tgb} final states the SUSY Sudakov terms. 
One main conclusion is that
these terms do exist and are all of "subleading" SL (linear) kind. 
At one TeV, working in a relatively "light" SUSY
scenario, where the heaviest SUSY mass if of a few (two, three) 
hundred GeV, the numerical effects of SUSY Sudakov terms on the various
observables are relatively small (a few percent at most), 
while at three TeV they become definitely large. 
The next step has been the analysis of final scalar 
(sfermion or Higgs) pair production for both massless
and massive final states. This has been performed both at the 
one-loop level and at resummed  subleading order
accuracy \cite{scalar}. By comparison of the two calculations 
it can be concluded that, in the assumed "light" SUSY
scenario the two approximations are
practically indistinguishable at one TeV, where their effect is,
as in the fermion case, relatively small,  but become 
drastically different (and both large, well beyond a relative 
ten percent) in the higher (two, three TeV)
energy region. Within this approach one would thus conclude 
that at the LC extreme energies the MSSM can be safely
treated, to subleading logarithmic order accuracy, at the one 
loop level for what concerns fermion, sfermion,
scalar Higgs production.\par
This conclusion could be of immediate practical consequence. 
In fact, it has been remarked in Refs.\cite{tgb,scalar}    
that, for production of fermions and squarks of the third 
family and of charged Higgs bosons, the coefficients of
the SL electroweak SUSY logarithms of Yukawa origin only 
depend on a common SUSY scale $M_S$, by definition the
heaviest SUSY mass involved in the electroweak component of 
the process, and on the ratio of the two scalar vevs 
$\rm tan \beta = v_2/v_1$. To the extent that a subleading order 
approximation can be considered as a reliable
description of the variation with energy of the observables of 
the process, i.e. that the missing terms of the
expansion can be adequately described by a constant component, 
this has allowed to propose a determination of $\rm tan\beta$ 
based on a number of measurements of the observables 
at different energies (roughly, on measurements of
their slopes), whose main features have been already illustrated 
in Refs. \cite{tgb,scalar} in a qualitative way for fermion
production and scalar production separately.\par
The aim of this note is that of proposing  a more quantitative 
determination of $\rm tan \beta$ from a \underline{combined}
analysis of the slopes in energy of fermion, sfermion, charged 
Higgs production. This will be done working at the
one-loop level, in an energy region around (below) 1 TeV. With 
this purpose, and to try to be reasonably
self-consistent, we briefly recall the structure of the various Yukawa 
contributions in the production of pairs of fermions ($t, b$), 
sfermions ($\widetilde t_{L, R}$, 
$\widetilde b_{L, R}$) and charged Higgs in the MSSM. 
The complete expressions of the asymptotic 
contributions can be found in Refs.\cite{tgb,scalar}, 
and we do not reproduce them here. Starting from them it 
is relatively straightforward to derive the
quantities that are relevant for this note, which are given in 
the following list.\\

\section{Complete List of Yukawa Effects in Cross Sections and Asymmetries}

We parametrize the Yukawa effects in the physical observables that we are going 
to
analyze and summarize them by giving a complete list.

Let us denote by ${\cal O}_n$, the various cross sections for production of 
sfermions
($\widetilde{t}_{L, R}$, $\widetilde{b}_{L, R}$), charged Higgs bosons $H^\pm$ 
and 
third generation fermions ($t$ and $b$). 
For top and bottom production we also include three basic asymmetry 
observables (unpolarized forward-backward asymmetry $A_{FB}$,
Left-Right asymmetry for longitudinally polarized $e^{\pm}$ beams
$A_{LR}$ and its forward-backward asymmetry $A_{pol}$).
In the case of top production the average helicity $H_t$
as well as its forward-backward and Left-Right
asymmetries $H_{FB,t}$ and $H_{LR, t}$
should be measurable by studying the leading top decay mode $t\to W b$.
The definition of these observables, in particular of the asymmetries, is conventional and
can be found in full details in Appendix B of~\cite{top}.

For cross sections ${\cal O}\equiv \sigma$, we define the relative one loop SUSY effect
as the ratio 
\bq
\epsilon_n(q^2) = \frac{{\cal O}_n(q^2)-
{\cal O}^{\rm Born + SM }_n(q^2)}{{\cal O}^{\rm Born + SM}_n(q^2)} ,
\eq
where ``SM'' denotes all the one loop terms
that do not involve virtual SUSY partners (sfermions, gauginos and extra Higgs particles). 
For asymmetries, we consider instead the absolute SUSY effect defined as the 
difference
\bq
\epsilon_n(q^2) = {\cal O}_n(q^2)-
{\cal O}^{\rm Born + SM}_n(q^2).
\eq

At one loop, in the asymptotic regime, the shifts $\epsilon_n$ can be 
parametrized as 
\bq
\label{expansion}
\epsilon(q^2) = \frac{\alpha}{4\pi}\ F(\tan\beta) \ln\frac{q^2}{M_S^2} + G + 
{\cal O}\left(\frac{M^2}{q^2}\right) ,
\eq
where, as we wrote, $F$ is a simple function of $\tan\beta$ only. Its explicit expression must be 
determined by performing a Sudakov (logarithmic) expansion of the one loop calculation. The detailed
analysis can be found in~\cite{susylog,tgb,scalar} and here we collect the various results 
for convenience of the reader.

\vskip 0.5cm
\noindent{\bf Sfermion cross sections}

\bqa
F(\sigma_{\widetilde{t}_L}) &=& - \frac{1}{M_W^2 s_W^2}\left(m_t^2\cot^2\beta
+m_b^2\tan^2\beta\right), \\
F(\sigma_{\widetilde{t}_R}) &=& - \frac{2}{M_W^2 s_W^2}\ m_t^2\cot^2\beta,  \\
F(\sigma_{\widetilde{b}_L}) &=& F(\sigma_{\widetilde{t}_L}),  \\
F(\sigma_{\widetilde{b}_R}) &=& - \frac{2}{M_W^2 s_W^2}\ m_b^2 \tan^2\beta
\eqa

\vskip 1.5cm
\noindent{\bf Charged Higgs cross section}

\bq
F(\sigma_{H^\pm}) = -\frac{3}{M_W^2 s_W^2}\left(m_t^2 \cot^2\beta
+m_b^2\tan^2\beta\right) .
\eq

\vskip 0.5cm
\noindent{\bf Fermion cross section}

\bqa
F(\sigma_t)  &=& \frac{1}{s^2_W M_W^2}\frac{1}{9-12 s_W^2+88 s_W^4} \times  \\
&& \left(
-3m_t^2 (3-4s_W^2+56s_W^4)\cot^2\beta-m_b^2 (9-12 s_W^2+8 
s_W^4)\tan^2\beta\right)  \nonumber \\
F(\sigma_b) &=& \frac{1}{s^2_W M_W^2}\frac{1}{9-24 s_W^2+ 40 s_W^4} \times  \\
&& \left(
-m_t^2 (9-24s_W^2+20 s_W^4)\cot^2\beta-3 m_b^2 (3-8 s_W^2+20 
s_W^4)\tan^2\beta\right) \nonumber
\eqa

\vskip 0.5cm
\noindent{\bf Fermion cross section asymmetries}\\

\bqa
F(A_{FB,t}) &=&  \frac{1}{M_W^2} \frac{72 s_W^2(3-4 s_W^2 -4 s_W^4)}{(9-12 
s_W^2+88 s_W^4)^2}
(m_t^2\cot^2\beta-m_b^2\tan^2\beta)  \\
F(A_{LR,t}) &=& \frac{1}{M_W^2} \frac{384 s_W^2(3-4 s_W^2 + s_W^4)}{(9-12 
s_W^2+88 s_W^4)^2}
(m_t^2\cot^2\beta-m_b^2\tan^2\beta)  \\
F(A_{pol,t}) &=& \frac{1}{M_W^2} \frac{120 s_W^2(9-12 s_W^2 + 8 s_W^4)}{(9-12 
s_W^2+88 s_W^4)^2}
(m_t^2\cot^2\beta-m_b^2\tan^2\beta)  \\
F(A_{FB,b}) &=&  \frac{1}{M_W^2} \frac{-18 s_W^2(3-8 s_W^2)}{(9-24 s_W^2+40 
s_W^4)^2}
(m_t^2\cot^2\beta-m_b^2\tan^2\beta)  \\
F(A_{LR,b}) &=& \frac{1}{M_W^2} \frac{-96 s_W^2(3-8 s_W^2 + 5 s_W^4)}{(9-24 
s_W^2+40 s_W^4)^2}
(m_t^2\cot^2\beta-m_b^2\tan^2\beta)  \\
F(A_{pol,b}) &=& \frac{1}{M_W^2} \frac{-30 s_W^2(9-24 s_W^2 + 20 s_W^4)}{(9-24 
s_W^2+40 s_W^4)^2}
(m_t^2\cot^2\beta-m_b^2\tan^2\beta)  
\eqa

\vskip 0.5cm
\noindent{\bf Fermion helicity and its asymmetries}\\
It has been shown in \cite{tgb} that the logarithmic parts of
these observables are related to those of the cross section
asymmetries as follows:

\bqa
F(H_f) &=& -\frac 4 3 A_{pol, f} \\
F(H_{FB,f}) &=& -\frac 3 4 A_{LR, f} \\
F(H_{LR,f}) &=& -\frac 4 3 A_{FB, f}
\eqa

\section{Limits and Confidence Regions for tan $\beta$}

The constant $G$ in Eq.~(\ref{expansion}) is a sub-subleading 
correction that does not increase with $q^2$ and depends on all 
mass ratios of virtual particles.
The omitted terms in Eq.~(\ref{expansion}) vanish in the high 
energy limit \cite{HH}.

To eliminate $G$ we assume 
that a set of $N$ independent measurements is available 
at c.m. energies $\sqrt{q_1^2}, \sqrt{q_2^2}, \dots,\sqrt{q_N^2}$
and take the difference of each measurement 
with respect to the one at lowest energy. For each observable, the resulting 
quantities
\bq
\delta_i = \epsilon(q_i^2)-\epsilon(q_1^2) ,
\eq
do not contain the constant term $G$ and take the simple form
\bq
\delta_i = F(\tan\beta^*) \ \ln\frac{q_i^2}{q_1^2} ,
\eq
where $\tan\beta^*$ is the {\em true} unknown value that describes the
experimental measurements. 

We now turn to a description of a possible strategy for the determination of 
$\tan\beta$.
It results from a non linear analysis of data that must deal with extreme 
situations 
where $\tan\beta$ is determined with a still reasonable but rather large 
relative error.

Let us label the various observables by the index $n=1,\dots , N_{\cal O}$ and 
denote by $\sigma_n(q^2)$ the experimental error on $\epsilon_n(q^2)$.
For each set of explicit measurements $\{\delta_n(q_i^2)\}$, 
the best estimate for $\tan\beta$ is the value that minimizes 
the $\chi^2$ sum
\bq
\chi^2(\tan\beta) = \sum_{i=1}^N\sum_{n=1}^{N_{\cal O}}
\frac{[F_n(\tan\beta) 
\ln\frac{q^2_{i+1}}{q^2_1} - \delta_{n, i}]^2}{4\sigma_{n, i}^2} ,
\eq
where $\delta_{n, i} \equiv \delta_n(q^2_i)$ and $\sigma_{n, i} \equiv 
\sigma_n(q_i^2)$. The factor $4$ in the above formula follows from the fact that we 
assume a conservative error $2\sigma_{n,i}$ on the difference $\delta_{n,i}$. In other words, 
we describe the experimentally measured quantity $\delta_{n, i}$ in terms of a 
normal Gaussian random variable distributed around the theoretical value 
computed at $\tan\beta^*$
\bq
\label{diff}
\delta_{n,i} = F_n(\tan\beta^*) \ln\frac{q^2_{i+1}}{q^2_1}  + 2\sigma_{n,i}\ 
\xi_{n,i} ,
\eq
with probability density for the independent fluctuations $\{\xi_{n,i}\}$ given 
by 
\bq
P(\{\xi_{n,i}\}) = \prod_{n,i} \frac{1}{\sqrt{2\pi}}\ e^{-\frac 1 2 \xi_{n,i}^2} .
\eq
In the following we shall simplify the analysis
by taking a constant $\sigma_{n,i}\equiv\sigma$ with typical values around 1\%. 
For each 
set of measurements we determine the optimal $\tan\beta$ that minimizes 
$\chi^2$. It is a function
of the actual measurements $\{\xi_{n,i}\}$ and the width of
its probability distribution $P(\tan\beta)$ determines the limits that can be 
assigned to the 
estimate of the unknown $\tan\beta^*$. 

The distribution $P(\tan\beta)$ cannot be computed analytically because of the 
highly 
non linear dependence of the MSSM effects on $\tan\beta$. However, it can be 
easily obtained
by Monte Carlo sampling. With this aim, we generate a large set of independent
realizations of the {\em measurements} $\{\xi_{n,i}\}$ and compute for each of 
them
$\tan\beta$. The histogram of the obtained values is a numerical estimate of the 
true $P(\tan\beta)$.

In previous papers~\cite{tgb,scalar}, we discussed a simplified approximate 
procedure 
and we determined the $1\sigma$ boundary on $\tan\beta$ by linearizing the 
dependence on 
$\tan\beta$ around the minimum of $\chi^2$. The bound that we derived is thus 
\bq
\label{linear}
\delta\tan\beta = 2\sigma\left(\sum_{n=1}^{N_{\cal O}} 
F_n'(\tan\beta^*)^2\right)^{-1/2}
\left(\sum_i \ln^2\frac{q_{i+1}^2}{q_1^2} \right)^{-1/2} .
\eq
This result can be trusted if the experimental accuracy $\sigma$ is 
small enough to determine a 
region around the minimum of $\chi^2$ where deviations from 
linearity can be neglected. 
It gives anyhow a rough idea of the {\em easy} regions where 
a determination of $\tan\beta$
from virtual one loop MSSM effects is not difficult.

In a more realistic analysis, however, this approximation can be misleading and 
possibly too
much optimistic, especially for values of $\tan\beta$ around $15$ where the 
linearized analysis 
predicts typical relative errors around $50\%$. For this reasons, we 
pursue in this paper the complete Monte Carlo analysis of the allowed range of 
$\tan\beta$.

\section{Numerical Analysis}

We begin by considering the full set of 16 observables consisting in:
\begin{enumerate}
\item cross sections for sfermion production in the case of final
$\widetilde{t}_L$, $\widetilde{t}_R$, $\widetilde{b}_L$, $\widetilde{b}_L$;

\item cross section for charged Higgs production;

\item cross sections and 3 asymmetries (forward-backward, longitudinal and 
polarized) for 
top, bottom  production;

\item 3 top helicity distributions (again forward-backward, longitudinal and 
polarized).
\end{enumerate}

\begin{figure}[htb]
\vspace*{1cm}
\[
\epsfig{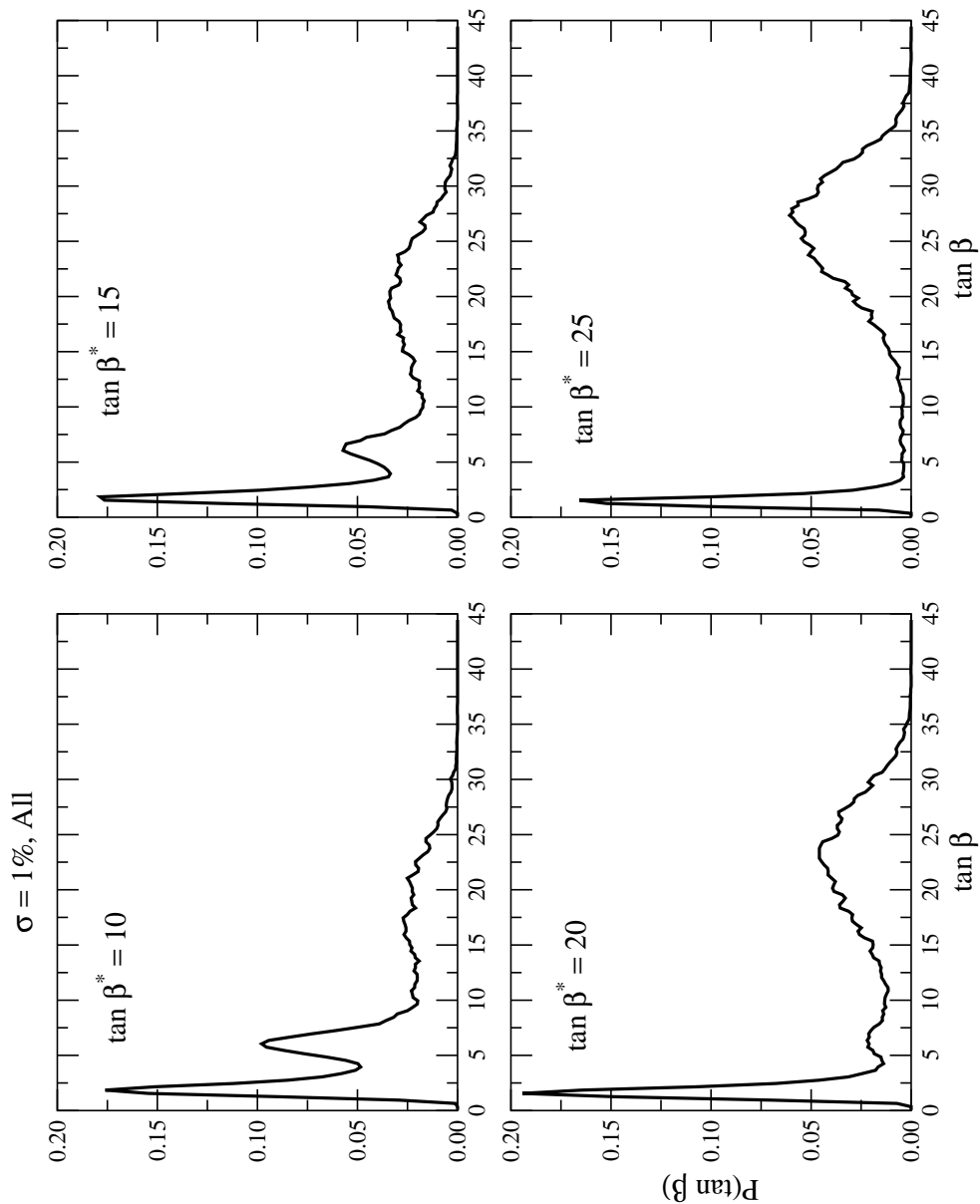}
\]
\vspace*{-0.5cm}
\caption[1]{
Histogram for the estimated $\tan\beta$ minimizing $\chi^2$. The experimental data 
included in the fit are all the observables discussed in the main text. We assume a precision 
$\sigma = 1\%$ on all data. The energy range is $0.6 < \sqrt{s} < 1.0$ TeV in this and the following
figures. The four boxes show what happens at the particular values 10, 15, 20 and 25 
of the parameter $\tan\beta^*$ that, we recall, is the true value.
}
\label{Histograms_0.010_All}
\end{figure}

We assume a set of $N=10$ measurements at
energies between 600 GeV and 1 TeV with an aimed experimental precision equal to 
1\% for all observables
at all energies. Within this ideal framework we have determined the probability 
distribution 
$P(\tan\beta)$ for $5 < \tan\beta^* < 40$. 
In Fig.~(\ref{Histograms_0.010_All}) we show the associated histograms in the 
four cases $\tan\beta^*=10, 15, 20, 25$. 
Even in the easiest case, $\tan\beta^*=25$, it is not possible to determine 
$\tan\beta$ in a reasonable way.
There is always a rather pronounced peak at small $\tan\beta$ in the histogram 
and the distribution is 
rather broad without a second peak recognizable around the exact value.

To analyze in a quantitative way these results, we compute from each histogram 
the 
standard deviation of the estimated $\tan\beta$. If the distribution can be 
characterized in terms
of a single dominant peak, then this is rough measure of its width. Of course, 
when the distribution is 
wide or when it is the sum of two large separated peaks, the standard deviation 
is a pessimistic 
estimate of the uncertainty on the parameter determination that can be improved 
by adding, for instance, 
some information excluding the regions corresponding to large (or small) values.

\begin{figure}[htb]
\vspace*{1cm}
\[
\epsfig{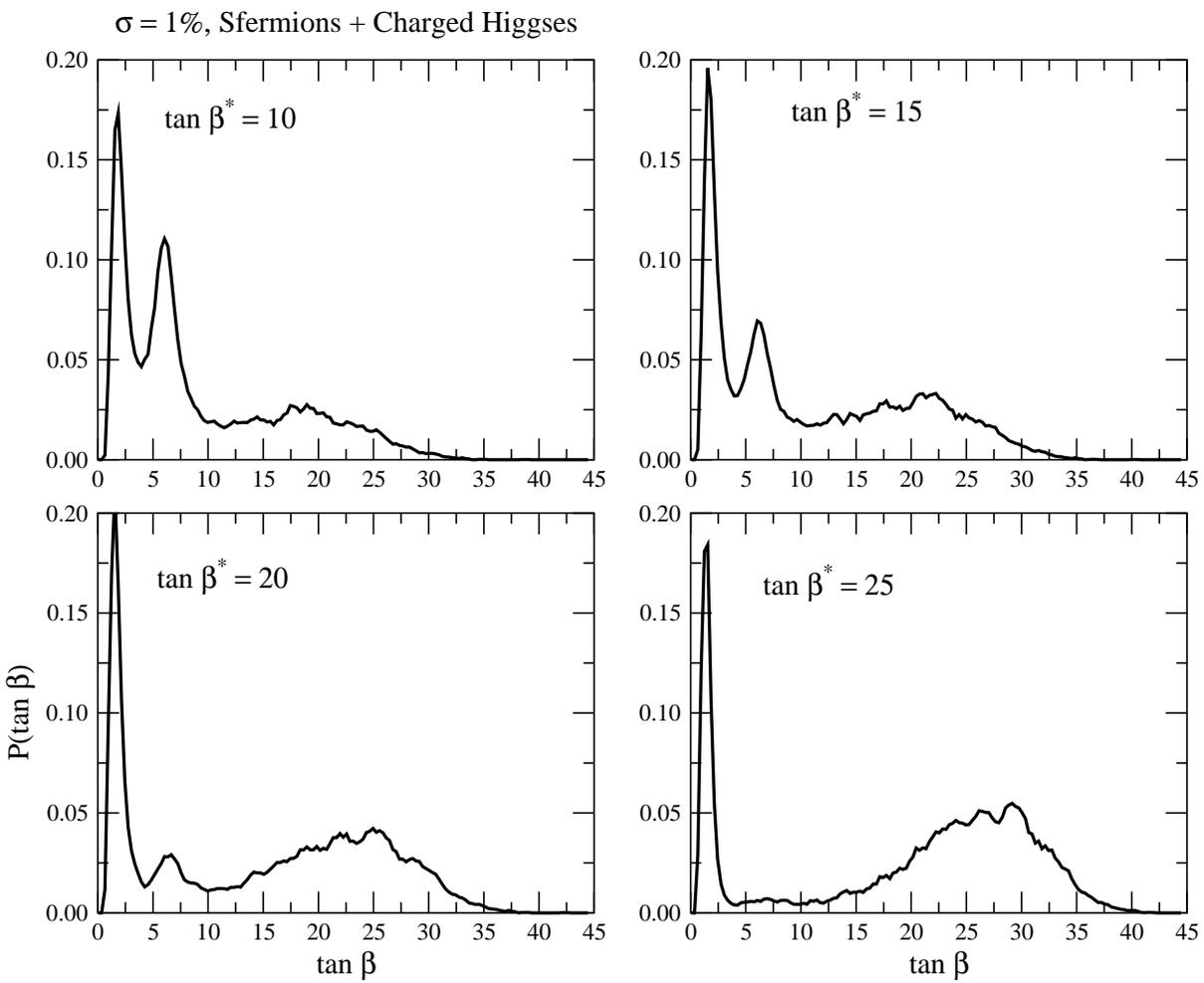}
\]
\vspace*{-0.5cm}
\caption[1]{
This figure is similar to Fig.~(\ref{Histograms_0.010_All}), but $\chi^2$
minimization is performed 
by considering only the observables associated to the production of sfermions and charged Higgses.
}
\label{Histograms_0.010_Sfermions_ChHiggses}
\end{figure}

\begin{figure}[htb]
\vspace*{1cm}
\[
\epsfig{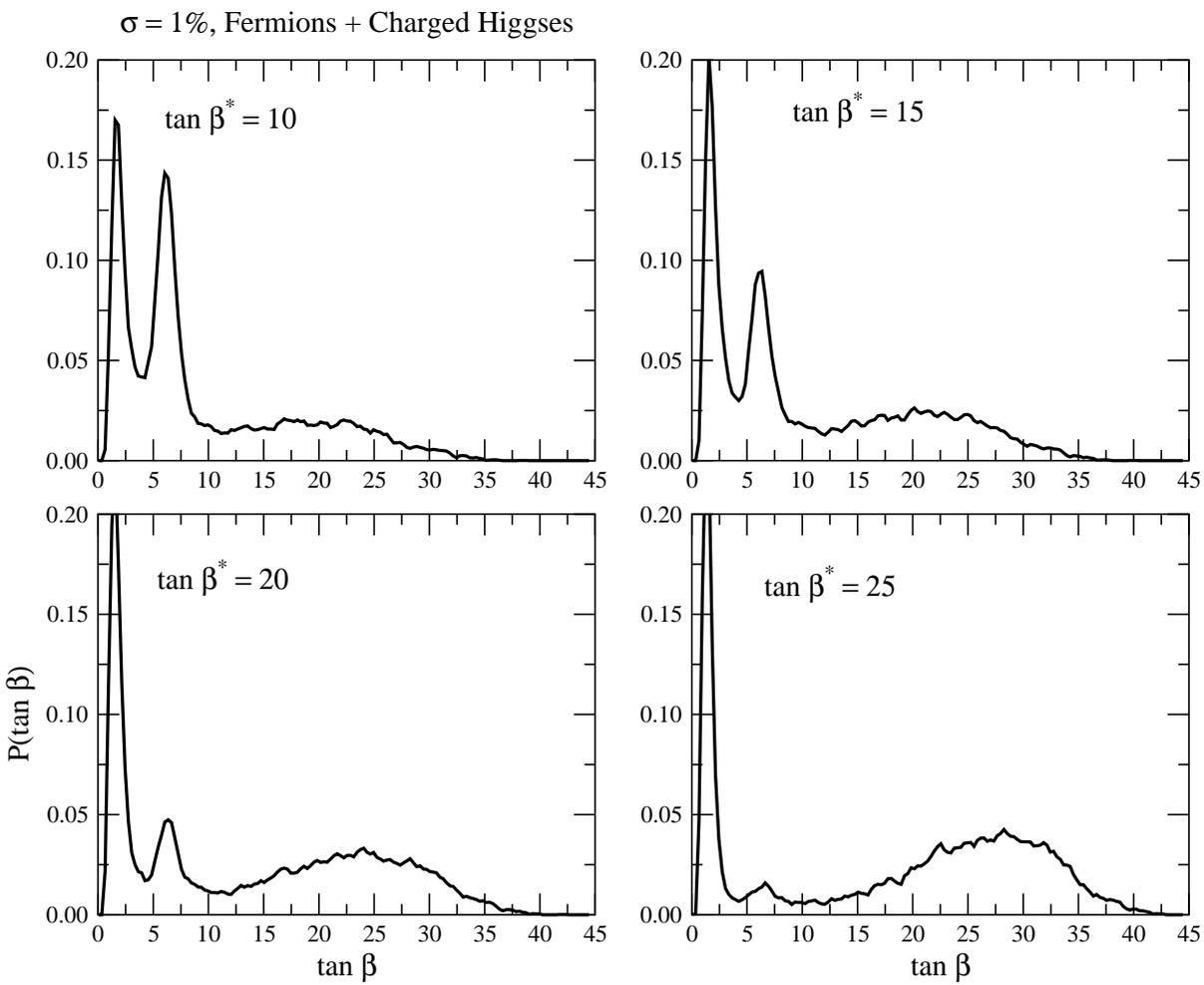}
\]
\vspace*{-0.5cm}
\caption[1]{
This figure is similar to Fig.~(\ref{Histograms_0.010_All}), but $\chi^2$
minimization is performed 
by considering only the observables associated to the production of heavy fermions (top and bottom)
and charged Higgses.
}
\label{Histograms_0.010_Fermions_ChHiggses}
\end{figure}

\begin{figure}[htb]
\vspace*{1cm}
\[
\epsfig{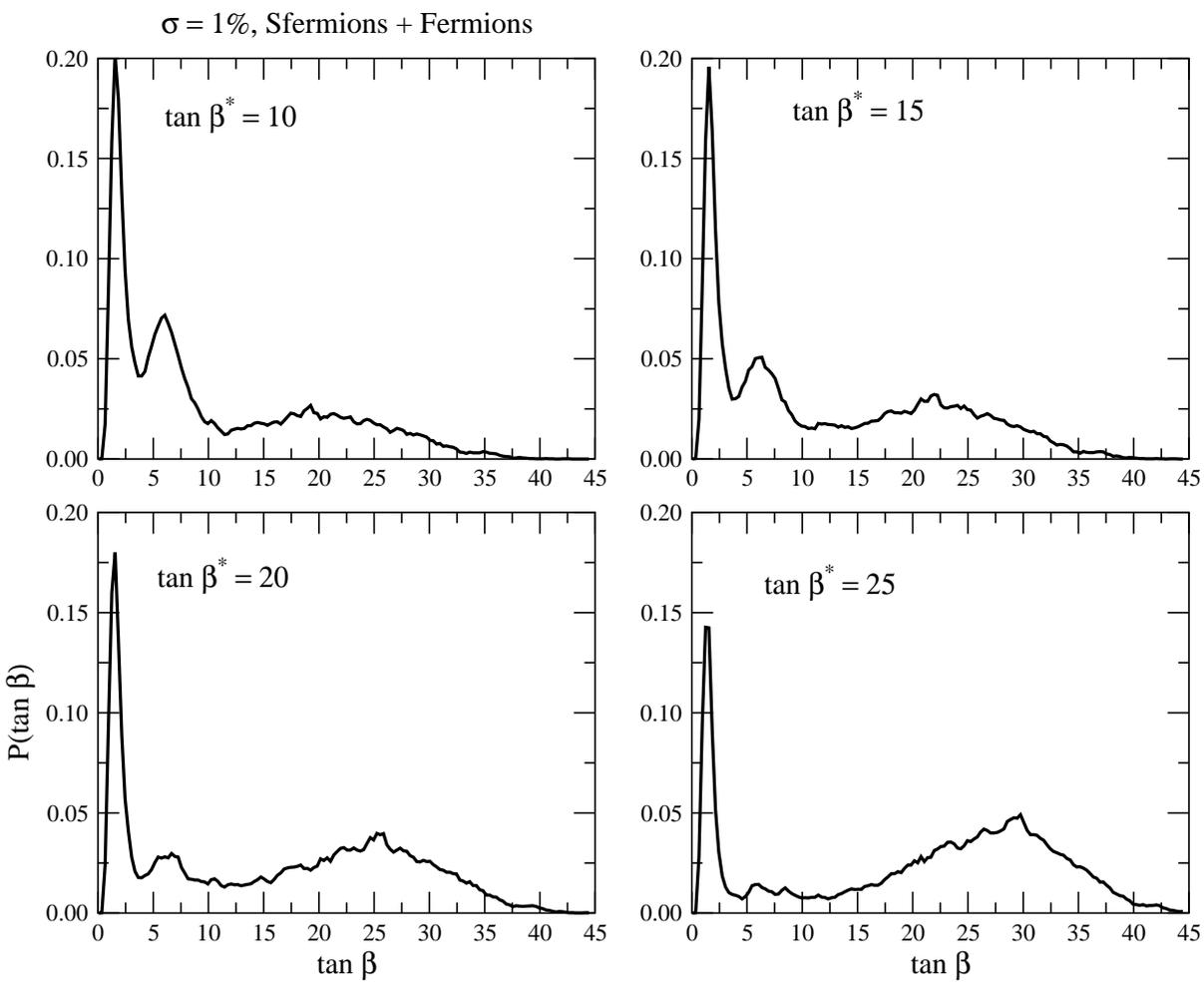}
\]
\vspace*{-0.5cm}
\caption[1]{
As a final case, this figure is again similar to Fig.~(\ref{Histograms_0.010_All}), but $\chi^2$
minimization is performed by considering only the observables associated to the production of 
heavy fermions (top and bottom) and sfermions.
}
\label{Histograms_0.010_Sfermions_Fermions}
\end{figure}

\begin{figure}[htb]
\vspace*{1cm}
\[
\epsfig{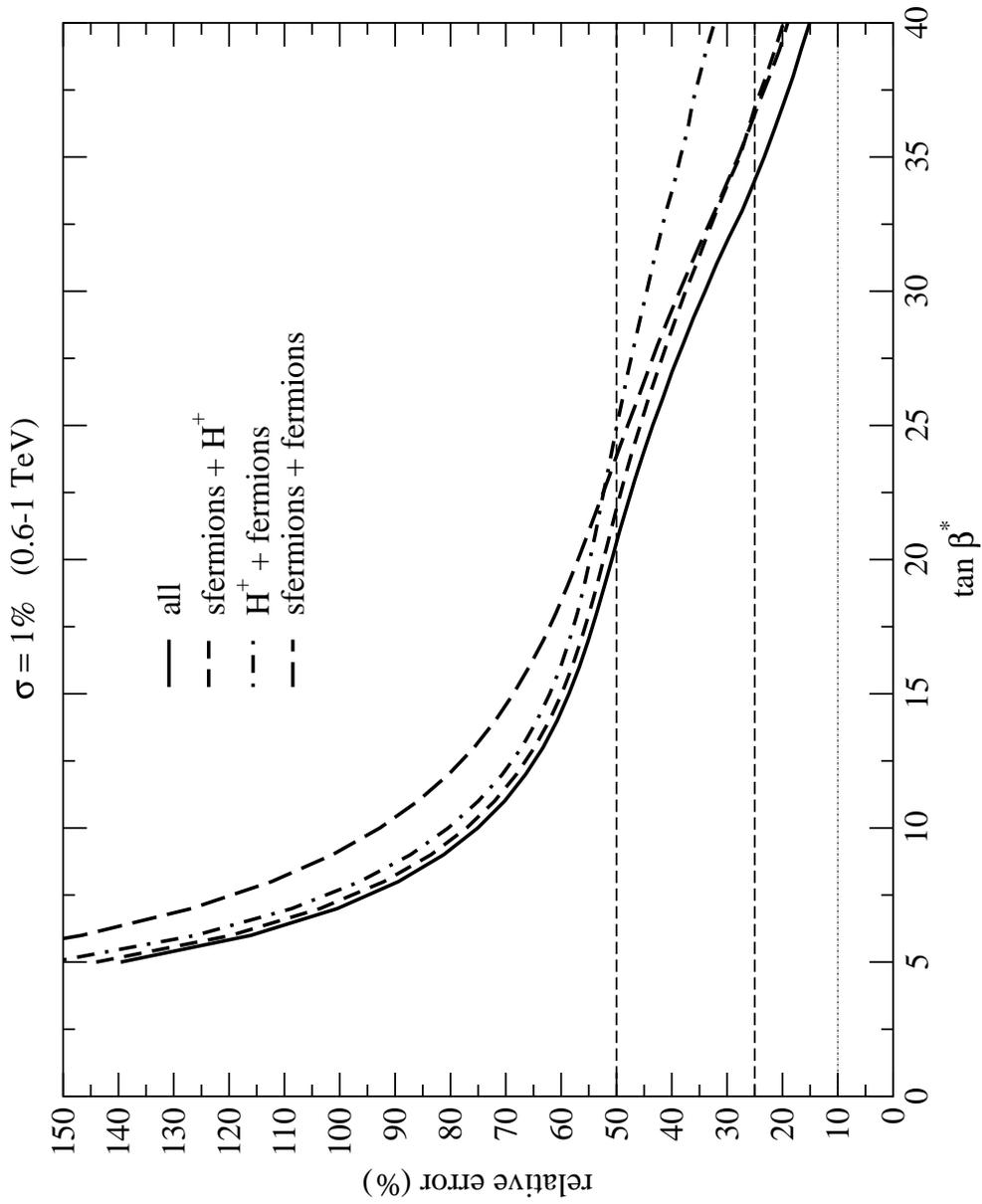}
\]
\vspace*{-0.5cm}
\caption[1]{
This figure shows the relative error in the estimate of $\tan\beta$ as a function of the
true unknown value $\tan\beta^*$. The four curves correspond to the different set of 
observables considered in the previous figures, as described in the legend.
}
\label{Error_Combined_0.010}
\end{figure}

The determination of $\tan\beta$ is almost completely driven by the observables 
related to 
sfermions and charged Higgses production. In 
Fig.~(\ref{Histograms_0.010_Sfermions_ChHiggses}),
we show the results obtained without observables related to top and bottom 
production.
In Fig.~(\ref{Histograms_0.010_Fermions_ChHiggses}), we show the results 
obtained with top, bottom and 
charged Higgs observables. The single measurement of charged Higgs cross section 
is not enough 
to determine $\tan\beta$ with this level of precision in the measures.
Finally, in Fig.~(\ref{Histograms_0.010_Sfermions_Fermions}), we show the 
results obtained with sfermions and fermions 
observables. The single measurement of charged Higgs cross section is not enough 
to determine $\tan\beta$ with this level of precision in the
measurements.
The plot of the relative error as a function of $\tan\beta$ for the various sets of 
observables (including the full case) is collected in Fig.~(\ref{Error_Combined_0.010}).

With these necessary
remarks in mind, we can analyze the standard deviation of the parameter 
histograms and the result is 
shown in Fig.~(\ref{Error_Combined_0.010}) (solid line). We see that for $\tan\beta< 20$ a 
determination with a relative error
smaller than $50\%$ is not possible.

\begin{figure}[htb]
\vspace*{1cm}
\[
\epsfig{file=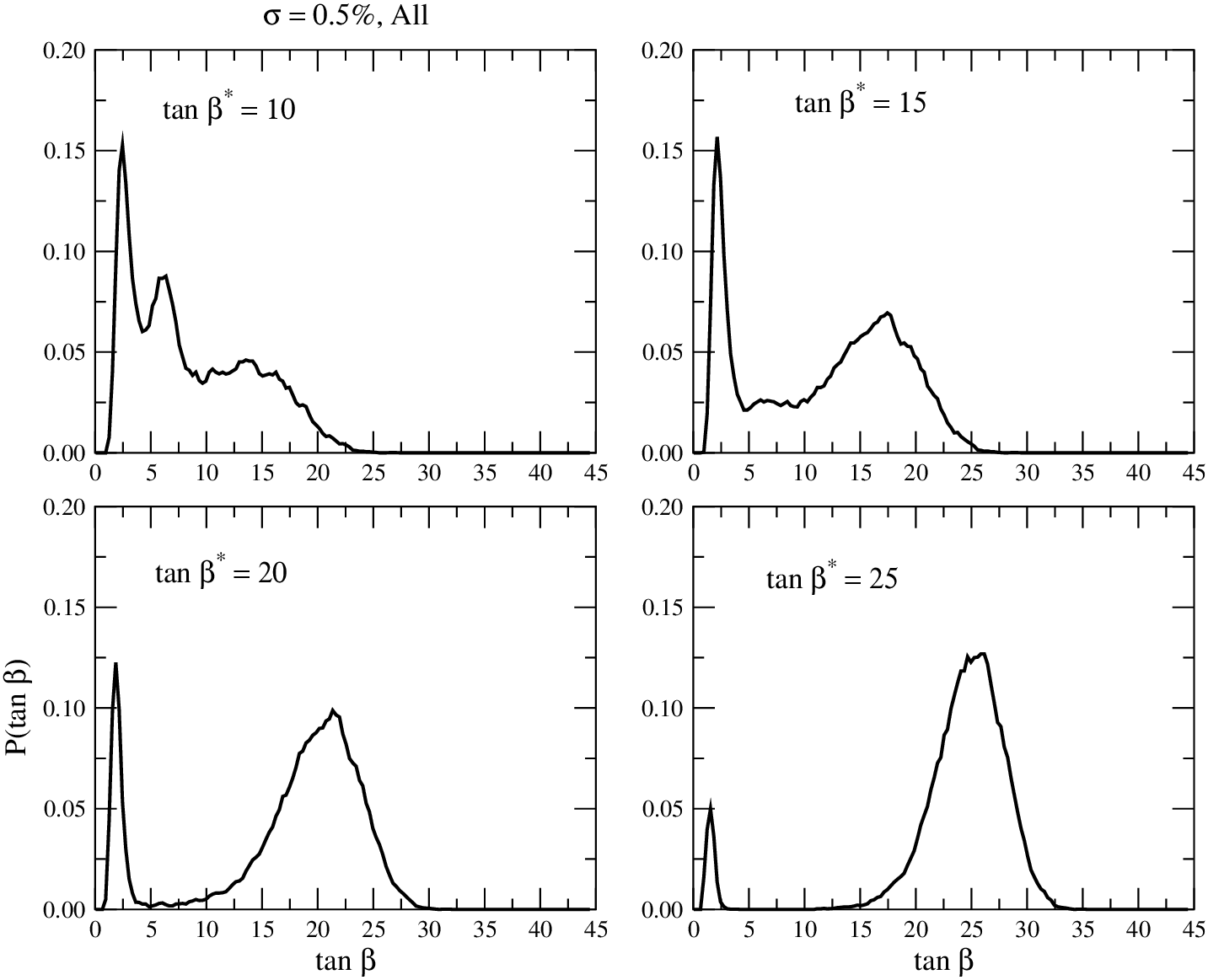,angle=90,height=16cm}
\]
\vspace*{-0.5cm}
\caption[1]{
The analysis of the results shown in this figure is precisely the same as in Fig.~(\ref{Histograms_0.010_All}), 
but with a global precision on the data reduced to $\sigma = 0.5\%$.
}
\label{Histograms_0.005_All}
\end{figure}

\begin{figure}[htb]
\vspace*{1cm}
\[
\epsfig{file=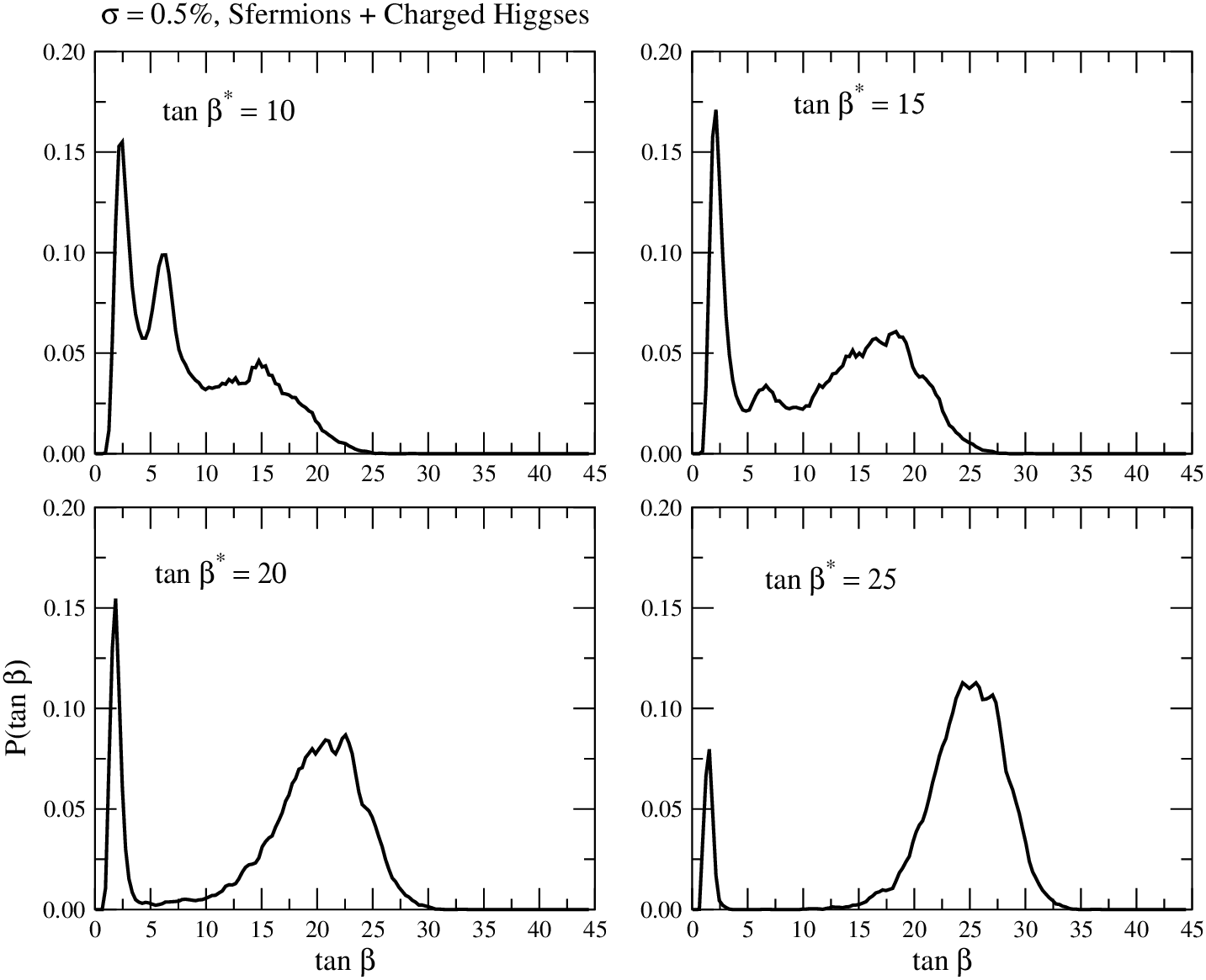,angle=90,height
=16cm}
\]
\vspace*{-0.5cm}
\caption[1]{
The analysis of the results shown in this figure is precisely the same as in 
Fig.~(\ref{Histograms_0.010_Sfermions_ChHiggses}), 
but with a global precision on the data reduced to $\sigma = 0.5\%$.
}
\label{Histograms_0.005_Sfermions_ChHiggses}
\end{figure}

\begin{figure}[htb]
\vspace*{1cm}
\[
\epsfig{file=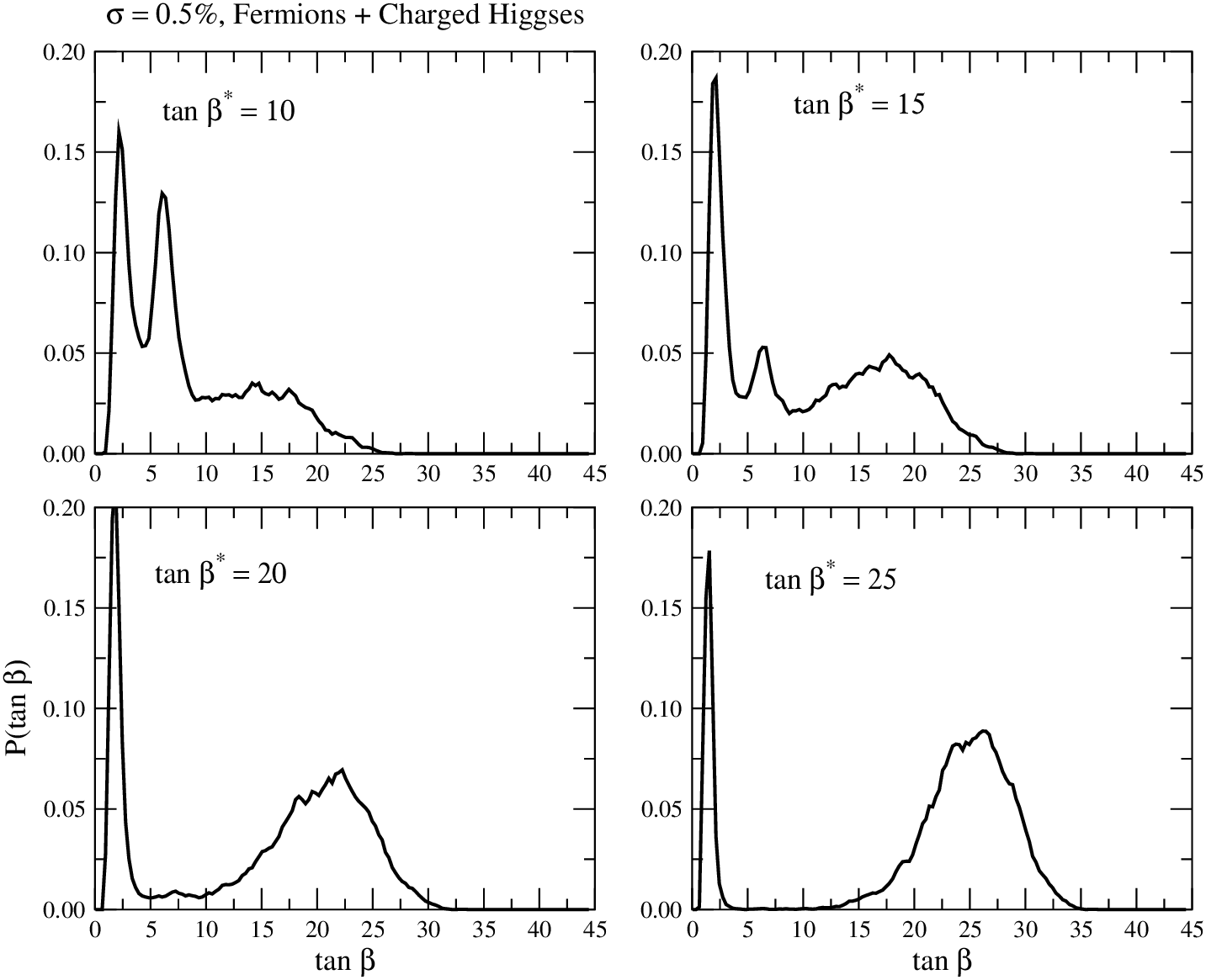,angle=90,height=
16cm}
\]
\vspace*{-0.5cm}
\caption[1]{
The analysis of the results shown in this figure is precisely the same as in 
Fig.~(\ref{Histograms_0.010_Fermions_ChHiggses}), 
but with a global precision on the data reduced to $\sigma = 0.5\%$.
}
\label{Histograms_0.005_Fermions_ChHiggses}
\end{figure}

\begin{figure}[htb]
\vspace*{1cm}
\[
\epsfig{file=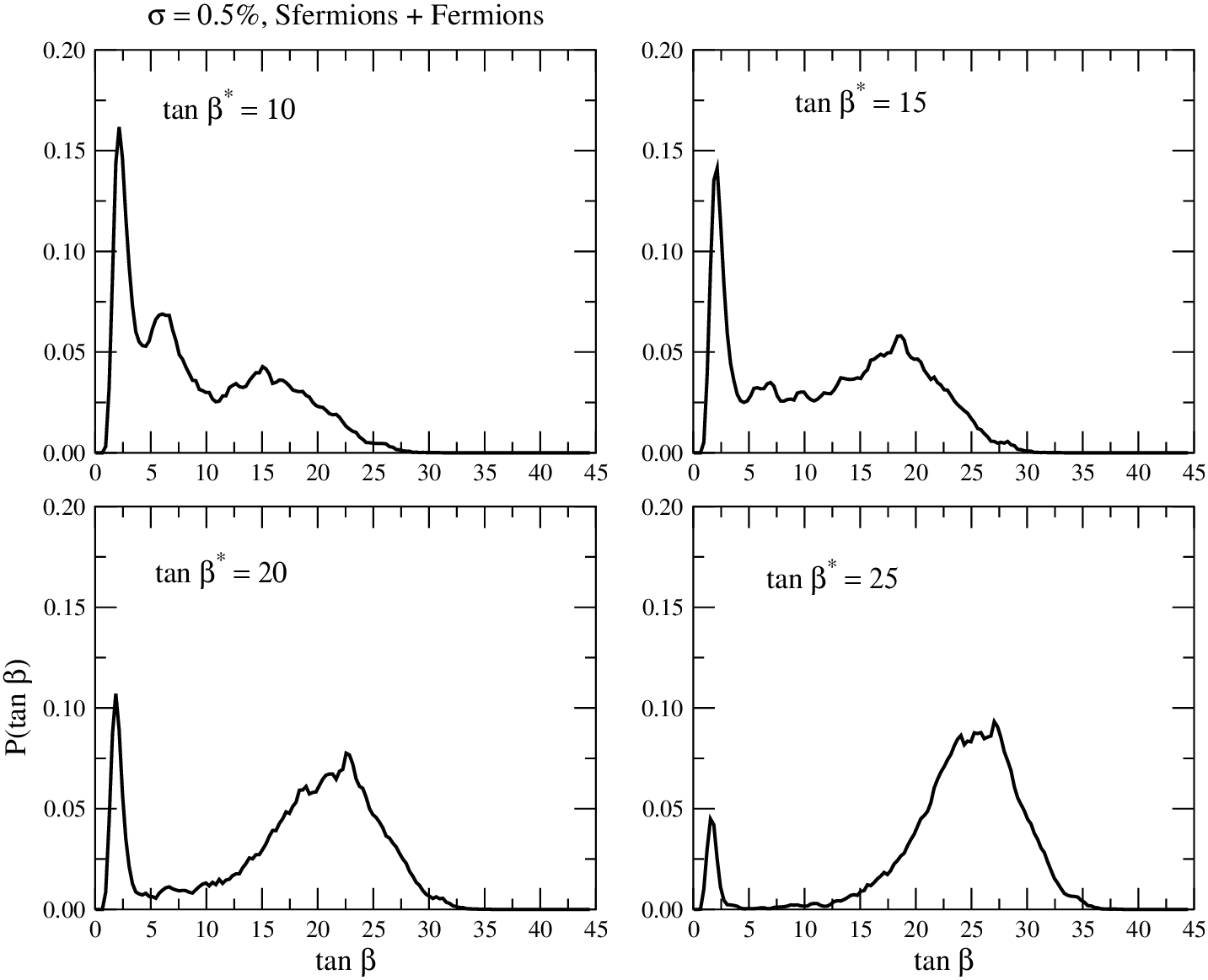,angle=90,height=
16cm}
\]
\vspace*{-0.5cm}
\caption[1]{
The analysis of the results shown in this figure is precisely the same as in 
Fig.~(\ref{Histograms_0.010_Sfermions_Fermions}), 
but with a global precision on the data reduced to $\sigma = 0.5\%$.
}
\label{Histograms_0.005_Sfermions_Fermions}
\end{figure}

\begin{figure}[htb]
\vspace*{1cm}
\[
\epsfig{file=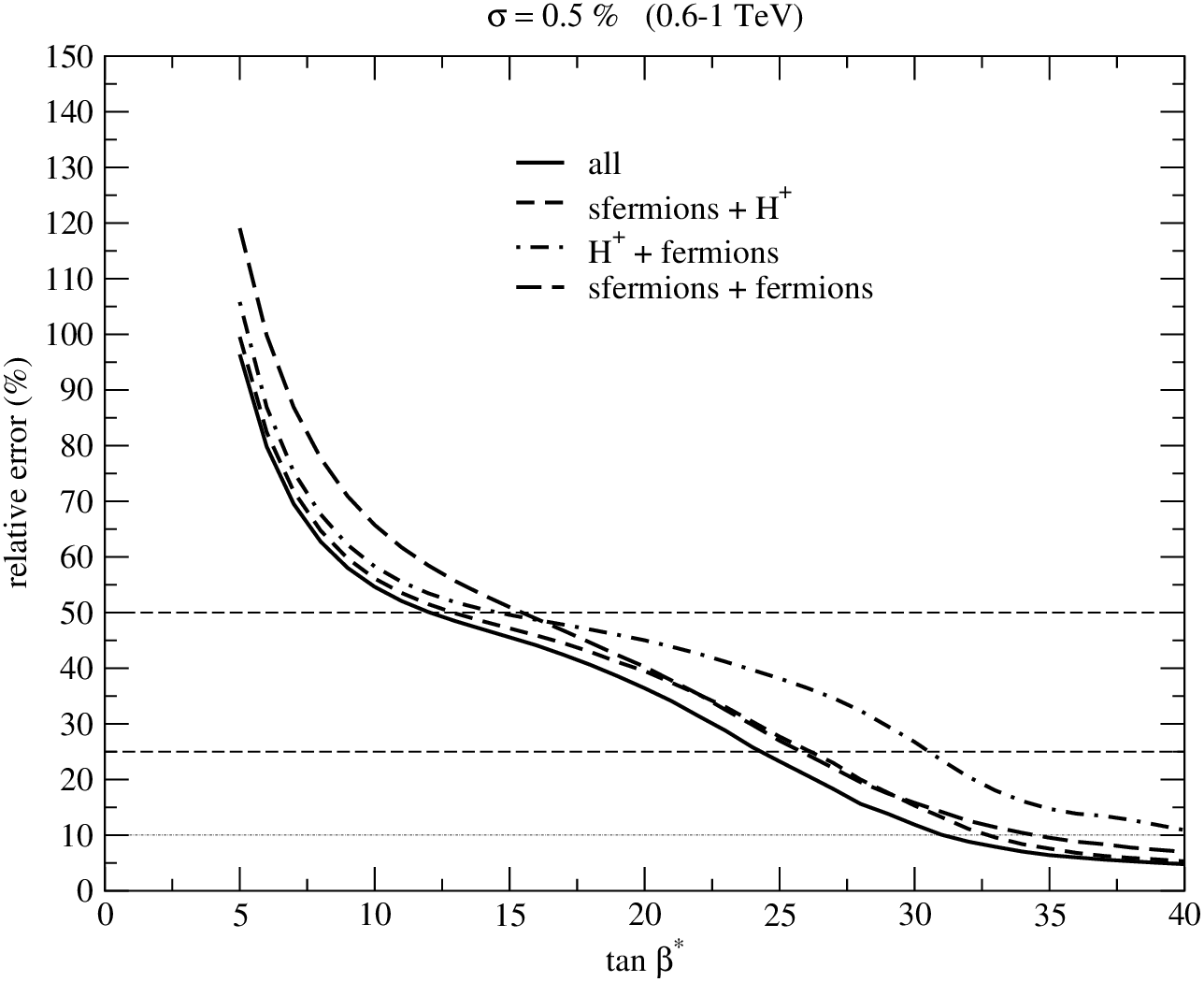,angle=90,height=16cm}
\]
\vspace*{-0.5cm}
\caption[1]{
The analysis of the results shown in this figure is precisely the same as in 
Fig.~(\ref{Error_Combined_0.010}),
but with a global precision on the data reduced to $\sigma = 0.5\%$.
}
\label{Error_Combined_0.005}
\end{figure}

If we consider still 10 measurements ranging from 600 GeV up to 1 TeV, but with 
a precision of
$0.5 \%$, then the scenario is quite better. 
In Fig.~(\ref{Histograms_0.005_All}), we see that for $\tan\beta^* > 20$, a well 
defined peak is visible 
in the rightmost part of the Figures roughly centered on the exact value. 
Fig.~(\ref{Error_Combined_0.005}) (solid line) allows to conclude that the relative error is 
smaller than $50\%$ for $\tan\beta^* > 13$ and 
smaller than $25\%$ for $\tan\beta^* > 25$.
Again, the role of the observables related to sparticle production is 
fundamental. 
In Fig.~(\ref{Histograms_0.005_Sfermions_ChHiggses})
we show what can be obtained without the information coming from 
top and bottom 
production. As one can
see, there is small difference with respect to the previous two 
figures.
Fig.~(\ref{Histograms_0.005_Fermions_ChHiggses}) shows the 
results obtained with 
top, bottom and 
charged Higgs observables.
Fig.~(\ref{Histograms_0.005_Sfermions_Fermions}) shows the 
results obtained with 
sfermions and fermions observables.
Again, Fig.~(\ref{Error_Combined_0.005}) collects the error as a function of $\tan\beta$ for 
the various considered cases.

As a final comment, we observe that a general feature of the histograms is the presence of a 
fake peak at small $\tan\beta$ as well at $\tan\beta\simeq 6$. The reason for this can be
understood by analyzing what happens by exploiting in the analysis just the (dominant) 
charged Higgs cross section as discussed in Appendix~\ref{app}.

\section{Conclusions}

In a "standard" SUSY model, all the gauge couplings are fixed 
and coincide with the corresponding SM
ones. For the couplings of the Yukawa sector, much more freedom 
is allowed. In the MSSM, one such
coupling is the ratio of the scalar vevs $\tan \beta$. We
have shown in this Note that, in a light SUSY scenario, a 
determination of $\tan \beta$ based on
measurements of the slope with energy of the combined set of 
observables of the 
three processes of fermion, sfermion, scalar charged Higgs
 production can
lead to a determination of this parameter with a relative 
error of 20-30 \% in a range of high values 
that would otherwise require difficult final state analyses
of Higgs decays, see the proposals in \cite{Datta-Gun}.\\

The main point of our approach is, in our opinion, 
the fact that in this
determination $\tan \beta$ is the only SUSY parameter to be measured:
all the other parameters give  vanishing contributions in the high 
energy limit. Isolating the various SUSY
parameters to be studied is in fact, in our opinion, a basic feature 
of any realistic "determination strategy".\\
      
In addition to the previous conclusion, we would like to add an
extra final comment. If a SUSY model
were different from the considered MSSM, in particular if it had 
a different Higgs structure (for example more Higgs doublets), 
the Yukawa couplings would be, quite generally, different. But the 
features of the Sudakov structure would remain essentially unchanged. 
This would lead to the possibility of deriving,
with minor changes in our approach, the components of the SUSY 
Yukawa sector that dominate the high energy behaviour in this model. 
In analogy with what
 was done at LEP1 for the "prediction",
from an analysis of one-loop effects, of the value of 
the top mass,  that remains in our opinion one of the biggest 
LEP1 achievements,  a combined set of high precision measurements
at future linear colliders physics could therefore produce a genuine
determination of this fundamental SUSY parameter. 

\appendix

\section{Analysis with the $H^\pm$ cross section alone}
\label{app}

It is interesting to analyze the role of charged Higgs production in $\tan\beta$ determination when no 
additional observables are exploited. In fact, at the level of precision we are working, some problems arise. 
To see why, let us denote $\tan\beta\equiv T$ and write $\chi^2$ 
explicitly:
\bq
\chi^2(T) = \sum_{i=1}^N
\frac{[(F(T)-F(T^*)) \ln\frac{q^2_{i+1}}{q^2_1} - 2\sigma \xi_i]^2}{4\sigma^2} ,
\eq
The function $F(T)$ is given by 
\bq
F(T) = -\frac{3\alpha}{4\pi M_W^2 s_W^2}(m_t^2\cot^2\beta+m_b^2\tan^2\beta)
\eq
If we denote $L_i = \log\frac{q_{i+1}^2}{q_1^2}$, the derivative of $\chi^2$ vanish when
\bq
F'(T)=0\ \Rightarrow \ T=\sqrt{m_t/m_b} \simeq 6.2 ,
\eq
and also at the solutions (if there are any) of 
\bq
F(T)-F(T^*) = 2\sigma\frac{\sum_i L_i \xi_i}{\sum_i L_i^2} ,
\eq
that we can write in a simpler way in terms of a new  normalized gaussian random variable 
$\widetilde\xi$:
\bq
\label{invert}
\Delta(T, T^*) = \widetilde\sigma\ \widetilde \xi
\eq
where
\bq
\Delta(T, T^*) \equiv F(T)-F(T^*),\qquad 
\widetilde\sigma = \frac{2\sigma}{(\sum_i L_i^2)^{1/2}} 
\eq
To discuss the solutions of Eq.~(\ref{invert}), we must consider the main features of 
$\Delta(T, T^*)$ for a given (unknown) $T^*$.
It tends to $-\infty$ for $T\to 0$ or  $\infty$ and vanishes at 
\bq
T_1 = T^*,\qquad T_2=\frac{m_t}{m_b}\frac{1}{T^*}
\eq
It is a convex function and attains its maximum value at $T=\sqrt{m_t/m_b}$ where 
\bq
\Delta_{max}(T^*) \equiv \Delta\left(\sqrt\frac{m_t}{m_b}, T^*\right) = 
\frac{3\alpha}{4\pi M_W^2 s_W^2}\left(\frac{m_t}{T^*}-m_b T^*\right)^2
\eq
Each would-be measurement corresponds to a value of $\widetilde\xi$ and to an associated random 
$T(\widetilde\xi)$ that is found by minimization of $\chi^2$. It is easy to see that two possibilities
can arise: 
\begin{enumerate}
\item [a)] $\widetilde\sigma\widetilde\xi \le \Delta_{max}(T^*)$: in this case, $\chi^2(T)$ has a double well
shape with a local maximum at $T=\sqrt{m_t/m_b}$ and two local minima 
around two points that are located around $T_1$ and $T_2$ and that tend  
toward them as $\widetilde\sigma\to 0$. 

\item [b)] $\widetilde\sigma\widetilde\xi > \Delta_{max}(T^*)$: in this case, $\chi^2(T)$ is concave
and has a global minimum at $T=\sqrt{m_t/m_b}$.
\end{enumerate}
If the would-be measurements are randomly generated, cases (a) and (b) will occur with a relative
frequency depending on $\widetilde\sigma$. 
For small $\sigma/\Delta_{max}$ the majority of cases will be (a) and we shall be able to identify 
two well defined peaks in the histogram of the reconstructed $T$. The first will be false 
and around $T_2$, the second will be true and around $T_1=T^*$. Of course, if several measurements
with independent dependencies on $T$ are combined, then it is possible to suppress the false peak.

If, on the other hand, $\sigma/\Delta_{max}$ is not small, then we shall fall in case (b) 
with very high probability and the reconstruction process will simply accumulate artificially 
at $T=\sqrt{m_t/m_b}$ just because Eq.~(\ref{invert}) has no solutions.

To give numerical values, with 10 measurements at $0.5\%$ between 600 GeV and 1 TeV, we find that 
the condition $\widetilde\sigma < \Delta_{max}$ forbids the analysis of the region $3 < \tan\beta < 13$
and in practice some other observable must be added (in the previous analysis we chose production of top or bottom).


\end{document}